# The role of cooling induced by mixing in the mass and energy cycles of the solar atmosphere

Andrew Hillier ⓘ,[1]★ Ben Snow ⓘ[1] and Iñigo Arregui[2,3]

[1]*Department of Mathematics, University of Exeter, Exeter EX4 4QF, UK*
[2]*Instituto de Astrofísica de Canarias, Vía Láctea s/n, E-38205 La Laguna, Tenerife, Spain*
[3]*Departamento de Astrofísica, Universidad de La Laguna, E-38206 La Laguna, Tenerife, Spain*

## ABSTRACT

In many astrophysical systems, mixing between cool and hot temperature gas/plasma through Kelvin–Helmholtz-instability-driven turbulence leads to the formation of an intermediate temperature phase with increased radiative losses that drive efficient cooling. The solar atmosphere is a potential site for this process to occur with interaction between either prominence or spicule material and the solar corona allowing the development of transition region material with enhanced radiative losses. In this paper, we derive a set of equations to model the evolution of such a mixing layer and make predictions for the mixing-driven cooling rate and the rate at which mixing can lead to the condensation of the coronal material. These theoretical predictions are benchmarked against 2.5D MHD simulations. Applying the theoretical scalings to prominence threads or fading spicules, we found that as a mixing layer grows on their boundaries this would lead to the creation of transition region material with a cooling time of ∼100 s, explaining the warm emission observed as prominence threads or spicules fade in cool spectral lines without the requirement for any heating. For quiescent prominences, dynamic condensation driven by the mixing process could restore ∼18 per cent of the mass lost from a prominence through downflows. Overall, this mechanism of thermal energy loss through radiative losses induced by mixing highlights the importance for considering dynamical interaction between material at different temperatures when trying to understand the thermodynamic evolution of the cool material in the solar corona.

## 1 INTRODUCTION

Observations show the existence of abundant cool, dense material in the hot, tenuous corona. This can be in the form of short-lived, dynamic chromospheric intrusions like spicules see panel a of Fig. 1 or, e.g. Beckers 1968, 1972; Sterling 2000; Bose et al. 2019; Hinode Review Team et al. 2019; Pereira 2019) and coronal rain (see panel a of Fig. 1 or, e.g. Hinode Review Team et al. 2019; Antolin 2020; Antolin & Froment 2022), or in the form of relatively long-lived structures like prominences (see both panels of Fig. 1 or, e.g. Labrosse et al. 2010; Parenti 2014). The close proximity of the cool and hot material implies a large temperature gradient, it would therefore be natural to expect thermal conduction to work to smooth out this gradient. However, the highly anisotropic nature of thermal conduction in the solar corona (where it predominantly transports heat along the magnetic field, Braginskii 1958) means that the cool plasma is to a degree thermally isolated from the surrounding corona by the magnetic field.

Even if the magnetic field can play a role to thermally isolate the cool material the corona, there is still a wide range of dynamic and thermodynamic evolution displayed by the cool material. For example, observations of quiescent prominences show that they possess downflows (Chae 2010) and rising plumes (e.g. Berger et al. 2010; Hillier 2018). Examples of both of these phenomena are shown in panel (b) of Fig. 1. It has even been observed that plumes can connect to the formation of downflows, potentially as a result of driving condensation of hot material (Berger et al. 2008). More strikingly some prominences show such persistent downflows, with the material falling like rain, that the downflows would drain the prominence mass in a few hours (Liu, Berger & Low 2012). Both spicules and prominence threads (shown in panel a of Fig. 1) have been observed to fade from the cool ($10^4$ K) spectral lines in which they are observed (e.g. Beckers 1972; de Pontieu et al. 2007; Anan et al. 2010; Sterling, Moore & DeForest 2010; Pereira et al. 2014; Okamoto et al. 2015; Rouppe van der Voort et al. 2015) and appear as emission in warm ($10^5$ K) spectral lines (Pereira et al. 2014; Okamoto et al. 2015; Rouppe van der Voort et al. 2015; Samanta et al. 2019).

Recent studies suggest that the dynamic motions of the cool material inside the corona, either through MHD waves (Terradas et al. 2008; Okamoto et al. 2015) or flows (Hillier & Polito 2018), can lead to shear-flow driven instabilities and then turbulent mixing (Magyar & Van Doorsselaere 2016; Hillier & Arregui 2019) and possibly turbulent heating (Antolin et al. 2015; Howson, De Moortel & Antolin 2017). The model of Hillier & Arregui (2019) explains how the mixing of cool and hot phases by the non-linear, turbulent stages of the Kelvin–Helmholtz (KH) instability leads to the formation of material at intermediate temperatures and densities. The temperature of this intermediate phase, approximately $10^5$ K (Hillier & Arregui 2019), coincides with the peak in the radiative loss function for the solar corona (e.g. Anzer & Heinzel 2000). Therefore, the mixed material will be out of thermal equilibrium and, as a result, cool. If this process is occurring in the solar atmosphere, it may play a relevant role in the thermodynamic evolution of plasmas in prominences, their fine structure, and in chromospheric spicules. The study of cooling-through-mixing may help us understand the quiescent prominence mass cycle with many downflows and the fading/appearance of spicules/prominence threads in different spectral lines.

★ E-mail: a.s.hillier@exeter.ac.uk

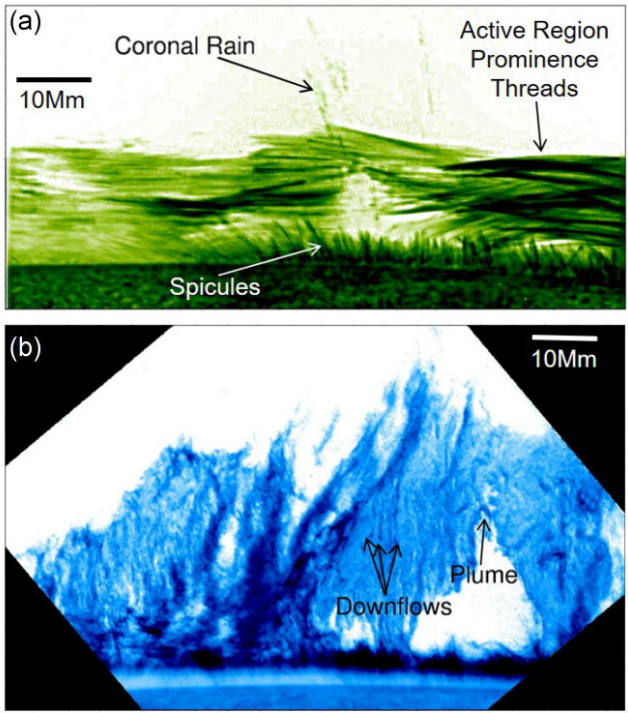

**Figure 1.** Images of prominences, coronal rain, and spicules taken using Hinode SOT. Panel (a) shows an active region prominence, with examples of coronal rain and spicules. Panel (b) shows a quiescent prominence presenting both downflows and upflows. Adapted from fig. 15 of Hinode Review Team et al. (2019).

The ideas in the paragraph above connect directly with the studies of radiative mixing layers in many different astrophysical systems. There has been many theoretical studies of this process in the interstellar and intracluster medium (e.g. Begelman & Fabian 1990), and in the circumgalactic medium (e.g. Ji, Oh & Masterson 2019; Fielding et al. 2020). The common physics being that mixing creates an intermediate temperature phase that can efficiently cool. As proposed by Hillier & Arregui (2019), this process may also occur in the solar atmosphere, and in this paper, we aim to explore those consequences. We derive a simple set of equations that can be used to understand the thermodynamic evolution of a system where mixing is driving cooling which we then apply to understanding the consequences on the thermodynamic evolution of cool material embedded in the hot corona.

## 2 THE MIXING-INDUCED COOLING MODEL

The model proposed here simply looks at the mixing of material at two different densities (and with that temperatures) to understand how the two different phases of material mix. Thermally, we assume that both phases of material that exist pre-mixing have inefficient radiative losses, meaning the cooling of these unmixed phases occurs on time-scales far longer than the mixing time-scale. As such these losses can be treated as negligible. However, with mixing creating material at a wide range of intermediate temperatures and densities (Hillier & Arregui 2019), we can consider a situation where the radiative losses in this intermediate phase can lead to efficient cooling. That is to say the only radiative losses that happen on a dynamic time-scale occur as a consequence of mixing of the two phases. In this situation, we propose that the mixing time-scale will control the dynamic evolution (and with that the creation of material out of thermal equilibrium), so this time-scale will determine the rate at which material cools, i.e. the bottleneck in the process is the mixing rate, and not the cooling time of the intermediate phase.

Following and extending Hillier & Arregui (2019), we use a model of a simple, discontinuous shear flow where the density and temperature vary across the jump. Arbitrarily assuming variation in the $y$ direction, we take a flow in the $x$ direction of $V_1$ for $y < 0$ and $V_2$ for $y > 0$ with the magnitude of the difference in these velocities being given by $\Delta V = |V_1 - V_2|$. Similarly, the densities in these two regions are $\rho_1$ and $\rho_2$ and the pressures in these two regions are $P_1$ and $P_2$ for $y < 0$ and $y > 0$, respectively. We focus on situations where the flow is approximately perpendicular to the magnetic field, i.e. the magnetic field is predominantly in the $z$ direction. We assume that the medium has a low plasma $\beta$ to be representative of the solar corona. The model proposed aims to capture the key aspects of a turbulent layer developing as a result of the KH instability growing on the initial interface of the discontinuous flow. This in done through determining the characteristics of a turbulent layer of width $W$ with its edges at positions $y = Y_1$ for $y < 0$ and $y = Y_2$ for $y > 0$.

The width of this mixing layer $W$, including its temporal evolution, can be modelled as following a self-similar evolution (Winant & Browand 1974) with the width at any given time given by

$$W = C_{\mathrm{mix}} \Delta V t, \quad (1)$$

where $C_{\mathrm{mix}}$ is a mixing constant that depends on the density contrast (Baltzer & Livescu 2020). Taking the characteristic velocity driving the turbulence to be the rms velocity in equation (33) from Hillier & Arregui (2019), we can predict the density dependence of the layer evolution with $W$ scaling as

$$W = \frac{C_1}{\sqrt{2}} \frac{(\rho_1 \rho_2)^{1/4}}{\sqrt{\rho_1} + \sqrt{\rho_2}} \Delta V t, \quad (2)$$

where $C_1$ is a constant that needs to be determined.

Hydrodynamic experiments suggest some range for the value for $C_1$. To match the quasi-2D dynamics (Winant & Browand 1974), a value of $C_1 \approx 0.5$ would be appropriate (Brown & Roshko 1974). Alternatively, to match with the 3D simulations of Baltzer & Livescu (2020), where the turbulence is more 3D, we would need to take $C_1 \approx 0.3$. Note this uncertainty is less than a factor of 2. We will discuss in Section 5 how magnetic fields may favour one situation over another, for now we are introducing this range to give some feel for the error associated with this derivation.

The next step is to take the derivative of equation (2) with respect to time to show that the rate at which the layer expands with time is given by

$$\dot{W} = \frac{C_1}{\sqrt{2}} \frac{(\rho_1 \rho_2)^{1/4}}{\sqrt{\rho_1} + \sqrt{\rho_2}} \Delta V. \quad (3)$$

Naturally this is independent of time.

Another prediction of the model of Hillier & Arregui (2019) is the shift in position of the mixing layer towards the low density layer,

with larger density differences leading to larger shifts. Taking $Y_1$ and $Y_2$ as the edges of the mixing layer on the low and high density side we have that

$$\frac{Y_1}{W} = -\frac{\sqrt{\rho_2}}{\sqrt{\rho_1} + \sqrt{\rho_2}}, \tag{4}$$

and

$$\frac{Y_2}{W} = \frac{\sqrt{\rho_1}}{\sqrt{\rho_1} + \sqrt{\rho_2}}. \tag{5}$$

The derivations for these two quantities can be found in Hillier & Arregui (2019).

With knowing the extent of the mixing region, the rate at which material is brought into the mixing layer can be calculated. Here, it is useful to define the full extent of the mixing volume, by considering that it has some length $L_{MIX}$ and depth $D_{MIX}$ that are fixed in time along with its varying width. Therefore, the rate of change of mass of the mixing layer ($\dot{M}$) is given as

$$\dot{M} = (\rho_1|\dot{Y}_1| + \rho_2|\dot{Y}_2|)L_{MIX}D_{MIX} \tag{6}$$

$$= \sqrt{\rho_1\rho_2}L_{MIX}D_{MIX}\dot{W} \tag{7}$$

$$= \frac{C_1\sqrt{\rho_1\rho_2}}{\sqrt{2}}L_{MIX}D_{MIX}\frac{(\rho_1\rho_2)^{1/4}}{\sqrt{\rho_1} + \sqrt{\rho_2}}\Delta V. \tag{8}$$

For the low-density material, the total mass of material ($M_1$) at this density brought into a mixing layer is given by

$$\dot{M}_1 \approx \frac{C_1\rho_1}{\sqrt{2}}L_{MIX}D_{MIX}\frac{\rho_1^{1/4}\rho_2^{3/4}}{(\sqrt{\rho_1} + \sqrt{\rho_2})^2}\Delta V. \tag{9}$$

Similarly for the high density material, the total mass of this material ($M_2$) brought in to the mixing layer is given by

$$\dot{M}_2 \approx \frac{C_1\rho_2}{\sqrt{2}}L_{MIX}D_{MIX}\frac{\rho_1^{3/4}\rho_2^{1/4}}{(\sqrt{\rho_1} + \sqrt{\rho_2})^2}\Delta V. \tag{10}$$

It is also possible to calculate the rate at which internal energy ($E_I$) is brought into the layer as

$$\dot{E}_I \approx \frac{C_1}{\sqrt{2}}L_{MIX}D_{MIX}\frac{(\rho_1\rho_2)^{1/4}}{(\sqrt{\rho_1} + \sqrt{\rho_2})}\Delta V$$
$$\times \frac{1}{\gamma - 1}\left(P_1\frac{\sqrt{\rho_2}}{\sqrt{\rho_1} + \sqrt{\rho_2}} + P_2\frac{\sqrt{\rho_1}}{\sqrt{\rho_1} + \sqrt{\rho_2}}\right), \tag{11}$$

with $P_1$ and $P_2$ the pressures in the low and high density regions. If we take that $P_1 = P_2 = P$ (i.e. gas pressure is constant across the interface) then

$$\dot{E}_I \approx \frac{C_1}{\sqrt{2}}L_{MIX}D_{MIX}\frac{(\rho_1\rho_2)^{1/4}}{(\sqrt{\rho_1} + \sqrt{\rho_2})}\Delta V \frac{P}{\gamma - 1}. \tag{12}$$

With information on how much thermal energy is entering the mixing layer in time, we can then estimate how much of this is lost through radiative cooling. Here, we use the assumption that the time-scale for cooling the mixed material is much shorter than the dynamic time-scale associated with the mixing. This implies that mixing becomes the bottleneck in the system, and as such it controls the overall rate at which material can cool, which will allow for an upper limit for the cooling rate to be determined. To estimate the energy loss, we need to estimate both the characteristic temperature of the material created by mixing, which initiates the cooling process, and the new, lower temperature that it reaches before the cooling time-scale of this now cooler material becomes longer than the mixing time-scale. We take the cut-off temperature after which cooling is inefficient, can be denoted as $T_{CUT}$. Assuming that $T_{CUT}$ is close to $T_2$ (the temperature of the cool layer) then we can estimate the characteristic temperature of the material that cools by the characteristic temperature of a mixing layer without cooling $T_{MIX} = \sqrt{T_1 T_2}$. This implies that the energy loss rate can be approximated by

$$\dot{E}_{loss} \approx \frac{T_{MIX} - T_{CUT}}{T_{MIX}}\dot{E}_I. \tag{13}$$

Because of the assumption of mixing being the bottleneck in this process, the exact form of the radiative loss function is not important, just it leads to cooling time-scales that are shorter than the mixing time-scale. Note here that the cooling model proposed is not based on thermal instability, the idea is that mixing creates fluid that is completely out of thermal equilibrium and it has to cool until it (approximately) reaches an equilibrium state.

It is here that the assumption of small plasma $\beta$ (say $\beta < 0.1$) becomes important. As the mixing layer cools, the gas pressure in the layer will reduce. If there is not a significant magnetic pressure in the mixing layer, this will result in a pressure imbalance between inside and outside of the layer, forcing it to contract. As we are assuming a dominant magnetic pressure, this decouples the cooling process from the dynamics allowing this model based on self-similar evolution to be applied even for regimes where internal energy is efficiently lost from the mixing layer.

Here, we have presented a model for the self-similar evolution of the mixing layer. Using this model, we have then been able to make predictions for the rates at which mass and internal energy are added to the mixing layer. Following these, equation (13) can be used to estimate the rate at which internal energy is lost from the mixing layer, i.e. the energy loss through radiative losses as a consequence of mixing. To complement this, under the mixing-cooling paradigm, equation (9) can be used to estimate the rate at which mass is lost from the hot phase and added to the cool phase (or at least the mass of material below $T_{CUT}$). It is these last two results that will be most important for applying these ideas to understand the consequences of mixing-induced cooling in the solar atmosphere.

## 3 NUMERICAL VALIDATION

### 3.1 Simulation model

To provide some proof-of-concept of the mixing-induced cooling model presented above, we perform a set of 2.5D simulations of mixing through the KH instability, where we add cooling through optically thin radiative losses to the mixing layer to compare with our predictions. We run these simulations using the (P_IP) code (Hillier, Takasao & Nakamura 2016) solving the compressible, ideal MHD equations in non-dimensional, conservative form with an added energy loss term

$$\frac{\partial \rho}{\partial t} + \nabla \cdot (\rho\mathbf{v}) = 0, \tag{14}$$

$$\frac{\partial}{\partial t}(\rho\mathbf{v}) + \nabla \cdot \left(\rho\mathbf{v}\mathbf{v} + P\mathbf{I} - \mathbf{B}\mathbf{B} + \frac{\mathbf{B}^2}{2}\mathbf{I}\right) = 0, \tag{15}$$

$$\frac{\partial}{\partial t}\left(e + \frac{B^2}{2}\right) + \nabla \cdot [\mathbf{v}(e + P) - (\mathbf{v} \times \mathbf{B}) \times \mathbf{B}] = -\rho^2 L(T), \tag{16}$$

$$\frac{\partial \mathbf{B}}{\partial t} - \nabla \times (\mathbf{v} \times \mathbf{B}) = 0, \tag{17}$$

$$\nabla \cdot \mathbf{B} = 0, \tag{18}$$

$$e \equiv \frac{P}{\gamma-1} + \frac{1}{2}\rho v^2, \quad (19)$$

$$T = \frac{\gamma p}{\rho}, \quad (20)$$

which allows us to calculate the evolution of the primitive variables, i.e. the density ($\rho$), velocity field ($\mathbf{v} = [v_x, v_y, 0]^T$), pressure ($P$), and magnetic field ($\mathbf{B} = [0, 0, B_z]^T$), through the evolution of the relevant conserved quantities. Note that equation (20) is the ideal gas law in the non-dimensional units used (where the temperature $T$ equals the sound speed squared).

These equations are solved using a fourth-order central difference approximation for spatial derivatives (calculated on a uniform mesh) with a four-step Runge–Kutta integration. Here, we do not explicitly include resistive or viscous terms in the equations. Dissipation at some level is inherent in the simulation due to the finite grid size and the use of flux limiters to smooth sharp structures. However, as can be seen later in Section 3.2, the level of the dissipation is not significant. The term $L(T)$ in the energy equation gives the functional dependence of the heat loss terms on the temperature $T$. We take $\gamma = 5/3$.

The initial conditions for the simulations are of a uniform pressure ($P = 1/\gamma$), and a uniform magnetic field $\mathbf{B} = \sqrt{2/(\gamma\beta)}[0, 0, 1]^T$ with $\beta = 0.05$. We impose an initial density profile of

$$\rho = \begin{cases} 100 & \text{for } y > 0, \\ 1 & \text{for } y < 0. \end{cases} \quad (21)$$

These choices result in the sound speed ($c_s = \sqrt{\gamma P/\rho}$) of unity in the low density (high temperature) region. Therefore, using the arbitrary lengthscale of $L = 1$, this leads to the sound crossing time as the normalizing time unit of the simulation.

The initial flow is set to be in the $x$ direction of

$$v_x = \frac{\Delta V}{101} \begin{cases} 1 & \text{for } y > 0, \\ 100 & \text{for } y < 0, \end{cases} \quad (22)$$

where we set $\Delta V = 0.2$. We have chosen the particular values of 1 and 100 to place the system in the rest frame for the linear instability, though this choice is arbitrary and does not affect the dynamic evolution. We impose a random noise perturbation in the $v_y$ field to excite the instability across multiple scales allowing it to develop at the preferred scale of the instability as well as breaking the symmetry of the system. We impose a maximum magnitude of 0.005 to allow for a short linear phase of the dynamics before the non-linear evolution takes over.

For these calculations, we use a domain between $-1$ and $1$ in the $x$ direction and $-1.5$ and $0.5$ in the $y$ direction. We use 400 grid points in both the $x$ and $y$ directions which gives a grid size of $\Delta x = \Delta y = 0.005$. Note that due to the finite resolution of the simulation, we have set a transition layer between the two flow regions of half-width 0.003, which gives two grid points in the initial jump. This will become slightly smoothed by the flux-limiters used in the numerical scheme (Hillier et al. 2016), but it was set to be thin to reduce any radiative losses from this transition region before mixing dynamics could take hold. We use periodic $x$ boundaries and reflective $y$ boundaries. As this is a 2.5D simulation, it should correspond to the value of the constant $C_1$ in equation (2) of 0.5.

The applied loss function L(T) is designed to maximize the optically thin radiative losses at the intermediate temperature. We use a loss function ($L(T)$) given by

$$L(T) = \text{sech}^2\left(\log_{10}\left(\frac{T}{T_{\text{PEAK}}}\right)(0.04\pi)^{-1}\right)/\tau_{\text{RAD}}, \quad (23)$$

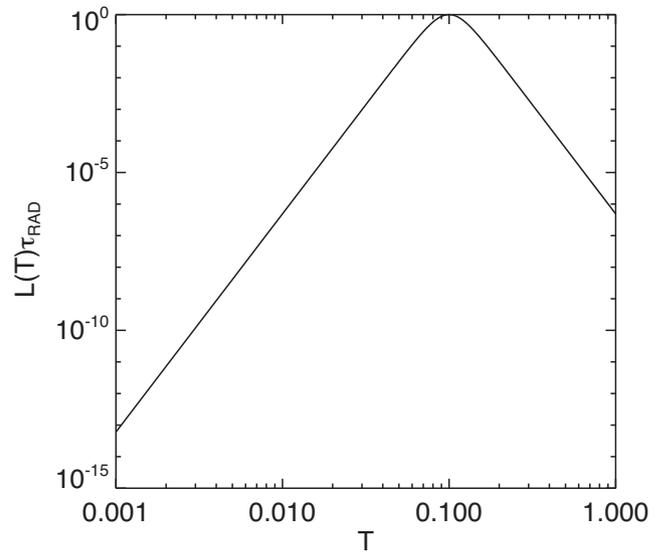

**Figure 2.** Plot of $L(T)\tau_{\text{RAD}}$ against $T$.

with $T_{\text{PEAK}}$ the temperature of the peak radiative losses (which we take to be 0.1 for our simulations) and $\tau_{\text{RAD}}$ the time-scale of the radiative losses (which we vary to investigate its effect). This form is chosen simply because it allows both the position of the peak radiative losses and the temperature range over which they are large to be controlled. We use this to mimic the optically thin radiative losses from the solar atmosphere where the intermediate temperature between cool and hot material in the solar atmosphere is at the peak in the radiative loss curve (e.g. Anzer & Heinzel 2000). As the analytical model presented in Section 2 should be applicable for any losses where the losses are efficient in the mixing layer but small in the pre-mixed phases, the exact form we chose is not critical for the process of benchmarking the model.

The physical meaning behind changing $\tau_{\text{RAD}}$ is that it changes the cooling time-scale with respect to the mixing time-scale allowing simulations where mixing dominates cooling or cooling dominates mixing to be investigated just through changing $\tau_{\text{RAD}}$. Fig. 2 shows the profile $L(T)\tau_{\text{RAD}}$ against $T$ that was used in these calculations.

### 3.2 Simulation results

Fig. 3 shows the density distribution for four cases: MHD ($\tau_{\text{RAD}} = \infty$), $\tau_{\text{RAD}} = 10^5$, $\tau_{\text{RAD}} = 10^3$, and $\tau_{\text{RAD}} = 10^1$ at time $t = 70$. Until this time, the system has first undergone a linear growth of the instability at small-scales focused on the region with a density and velocity jump. Following this, there has been a period of successive non-linear interactions resulting in the formation of a turbulent mixing layer that is increasing in size (e.g. Zhou et al. 2019).

A key point to note here is that though there are some differences in the density distributions shown in Fig. 3, the general properties that can be seen in the mixing layer (its width, its shift to the low density side, type of structures formed) have not changed in any drastic way through changing the cooling rate. This is a consequence of the low plasma $\beta$ which means that even as gas pressure is lost from the mixing layer, the magnetic pressure (which is the main pressure component) does not have to vary much to compensate. This means that the dynamics stay close to the self-similar evolution expected for the zero-cooling MHD simulation with flow and density profiles close to those predicted by Hillier & Arregui (2019).

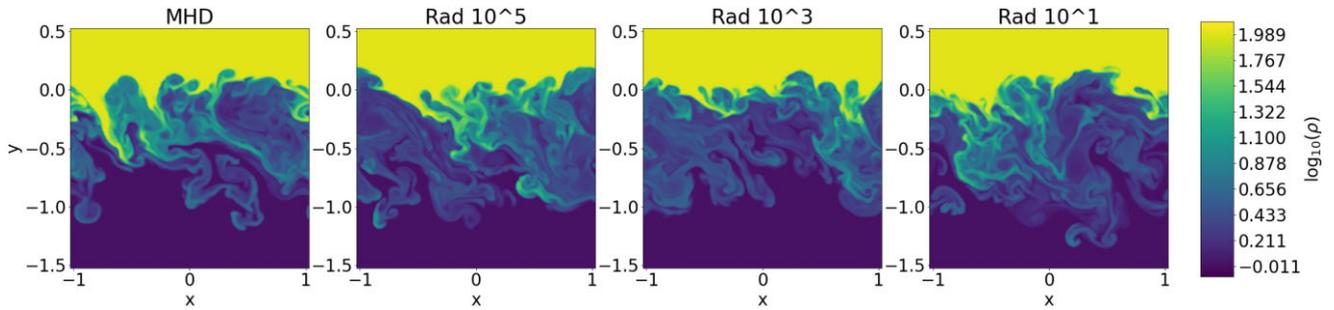

**Figure 3.** Density evolution for (left- to right-hand panel) the MHD (no cooling), weak cooling ($\tau_{RAD} = 10^5$), intermediate cooling ($\tau_{RAD} = 10^3$), and strong cooling ($\tau_{RAD} = 10^1$) cases.

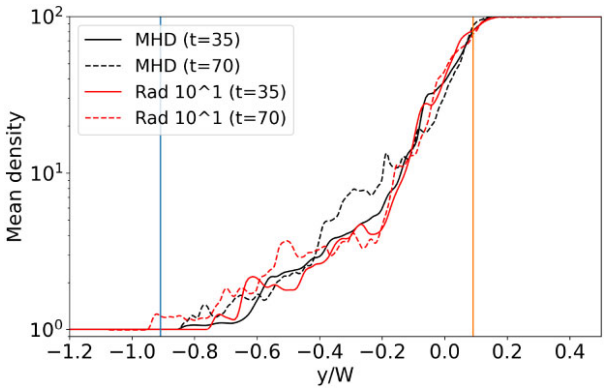

**Figure 4.** Plot of $x$ averaged density for the reference MHD simulation (black lines) and the $\tau_{RAD} = 10^1$ simulation (red lines) for $t = 35$ (dashed lines) and $t = 70$ (solid lines) against $y/W(t)$. Vertical lines show the predicted edges of the mixing layer from equations (4) (blue line) and (5) (orange line).

As the impression of Fig. 3 is that the inclusion of cooling is not making any vast changes to the mixing dynamics, we can investigate this further by comparing the predicted and actual sizes and shifts of the mixing layer. Fig. 4 shows the $x$ averaged density distribution for the reference MHD case and the $\tau_{RAD} = 10^1$ case at $t = 35$ and 70. The $x$ axis of the figure is $y/W$, which (assuming the theoretical predictions are correct) collapse the plots at different times onto a single curve. Considering the large-scale structure that can be seen in Fig. 3, it is not surprising there are differences, but the curves (both comparing between MHD and $\tau_{RAD} = 10^1$ cases and at different times) give similar distributions. The blue and orange vertical lines give the predictions (from equations 4 and 5) of the lower and upper bounds of the mixing layer. Again these can be seen as reasonably accurate predictions of the limits of the mixing layer. This provides some evidence that the scalings developed in the previous section will be informative for the thermodynamic evolution of the mixing layer.

Looking at the temperature distribution for the different simulations, see Fig. 5, there is clearly a large difference between the simulations. The reference MHD simulation and the weak cooling case ($\tau_{RAD} = 10^5$) display very similar temperature distributions. However, the case for $\tau_{RAD} = 10^1$ there is a clear change in the distribution. Almost all the material in the turbulent layer has cooled down to a low temperature of $\sim 10^{-1.45}$ or less. This clustering of material at or below this temperature can be seen clearly in the histogram in Fig. 6. Physically this can by understood as the temperature band where cooling is inefficient on the mixing time-scale so dynamically material can accumulate at these temperatures as mixing can create material at these temperatures faster than it can cool.

Fig. 7 shows the total internal energy with time for the MHD reference case, and a range of cooling times. For the reference MHD case and the longest cooling time used $\tau_{RAD} = 10^5$ there is evidence of a small level of heating present. However, for the strongest cooling simulations, there is a drastic reduction in the internal energy of the system as a result of cooling connected to the mixing. The black line is the slope predicted from equation (12) for the simulation parameters, which gives $T_{MIX} = 0.1$ and using Fig. 6 as a guide we take $T_{CUT} = 10^{-1.45}$. The prediction shown by the black line gives a reasonable approximation of the energy loss for the strong cooling cases.

We can also use the simulations to benchmark the rate at which the amount of cool material evolves with time. For a pure mixing simulation without radiative losses, it should be expected that cool material will be mixed with hot material to create the warm phase, resulting in a reduction over time of cool material. However, with radiative losses included the warm phase will cool, adding more material to the cool phase. Fig. 8 shows the evolution in time of the total material below $T_{CUT}$. In the MHD simulation (blue line), the role of mixing is to create intermediate phase material, so we can see the expected consistent decrease in the amount of material below the temperature with time. For the simulation with $\tau_{RAD} = 10^1$ (orange line), we find that the total mass below the cut-off temperature keeps increasing with time. The black line shows the simple prediction using the gradient determined from equation (9) for how the total mass below the cut-off temperature should grow with time. This gives an over prediction of the total mass held beneath these temperatures. However, the dashed black line is added to highlight that at late times, the amount of material below the cut-off temperature is growing at approximately the predicted rate.

## 4 APPLICATION TO THE SOLAR ATMOSPHERE

Having shown that we can use our model to give a reasonable prediction of the cooling rate and the rate at which mass is added to the cool phase, we can use these estimates to give a prediction for how mixing between cool material (e.g. prominence or spicule material) and the corona can lead to cooling. The consequences of this will be energy losses and an increase of mass of the cool material which we will explore in more detail here.

### 4.1 Replenishing prominence mass

Many quiescent prominences are observed to have downflows (see panel b of Fig. 1 or Chae 2010). These downflows can be so extensive that they would drain the whole mass of a prominence in just a few hours (Liu et al. 2012). However, these prominences with a huge number of persistent downflows exist in the corona for significantly longer than this time period (Liu et al. 2012), meaning there has to be a very efficient mechanism for replenishing the prominence mass. As these flows may be unstable to flow driven instabilities (e.g. Hillier & Polito 2018) and simulations have shown that these flows can occur perpendicular to the prominence magnetic field (Hillier et al. 2012; Kaneko & Yokoyama 2018), we investigate if a mixing-cooling process (for schematic see panel a of Fig. 9) can make a significant contribution to the mass budget of cool material in the prominence corona system by cooling coronal material as a consequence of mixing.

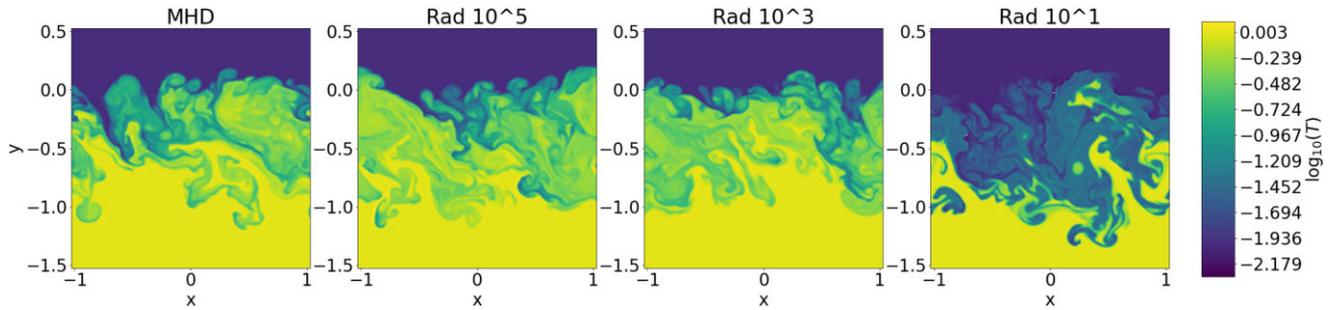

**Figure 5.** Temperature evolution for (left- to right-hand panel) the MHD (no cooling), weak cooling ($\tau_{RAD} = 10^5$), intermediate cooling ($\tau_{RAD} = 10^3$), and strong cooling ($\tau_{RAD} = 10^1$) cases.

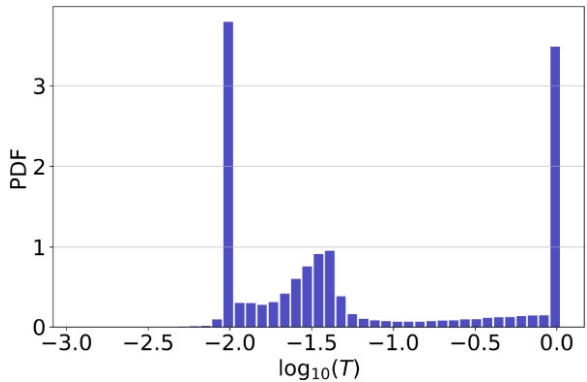

**Figure 6.** Histogram of the pdf for $\log(T)$ of the $\tau_{RAD} = 10^1$ case at $t = 70$. Cluster of material around $\log(T) = -1.45$ is where material in the mixing layer has cooled.

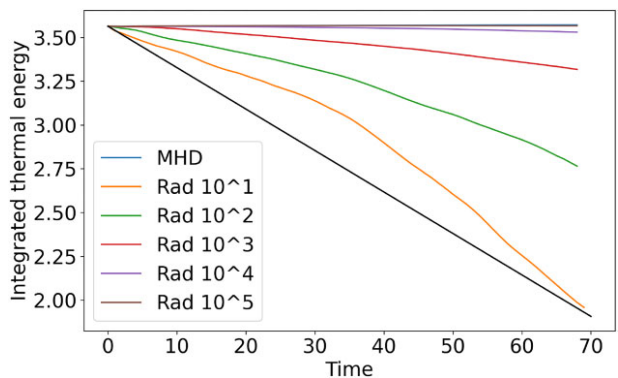

**Figure 7.** Change in total internal energy with time for different $\tau_{RAD}$ values. The reference MHD case is shown in blue, $\tau_{RAD} = 10^1$ in orange, $\tau_{RAD} = 10^2$ in green, $\tau_{RAD} = 10^3$ in red, $\tau_{RAD} = 10^4$ in purple, and $\tau_{RAD} = 10^5$ in brown. The black line shows the predicted slope for cooling rate based on cooling times being much shorter than mixing times and $T_{CUT} = 10^{-1.45}$ as is appropriate for the $\tau_{RAD} = 10^1$ simulation.

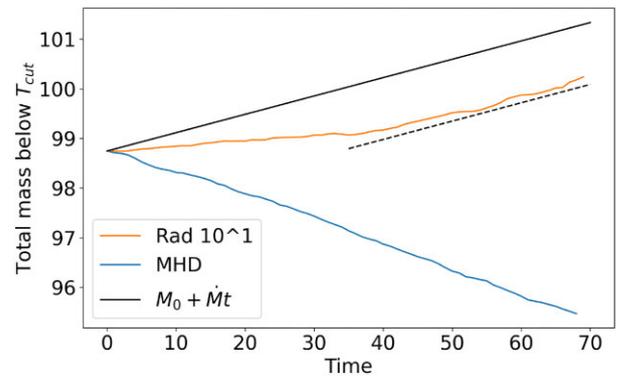

**Figure 8.** Evolution of the total mass below $T_{CUT}$ against time for the MHD case (orange line) and the $\tau_{RAD} = 10^1$ case (blue line). The solid black line is the predicted mass (with gradient determined from equation 9) if cooling starts at the beginning of the simulation. The dashed black line is the predicted slope of the mass increase used to highlight how the prediction matches the later time of the simulation.

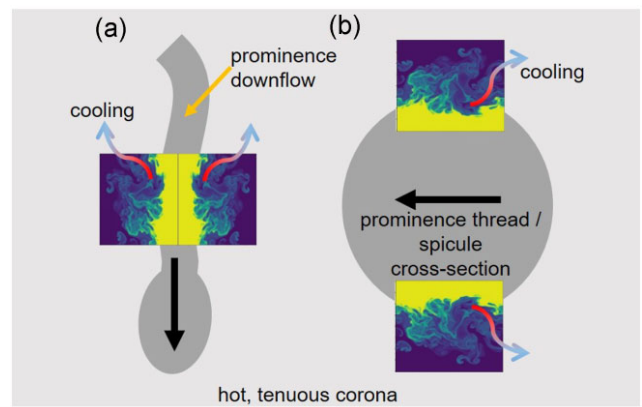

**Figure 9.** Schematic diagram showing the connection between the KH turbulence and (a) a prominence downflow with the black arrow showing the direction of flow and (b) the cross-section of an oscillating prominence thread or spicule with the black arrow showing the instantaneous direction of motion. Snapshots of the density structure from the $\tau_{RAD} = 10^1$ case are overlaid to give an impression of the structure of the small-scale turbulent mixing motions that we are modelling. Red-blue arrows from the mixing layer show the loss of thermal energy through photons.

To understand the rate at which these flows can add mass to a prominence, we can use equation (9), which (by assuming all the mass involved in mixing between the prominence and corona would cool to prominence temperatures) would give an estimate of the rate at which cool material is formed by a downflow. Here, we take that the downflowing thread has a height $H_{\rm THREAD}$, an observed width of $W_{\rm THREAD}$ and a depth along the line-of-sight $D_{\rm THREAD}$. With the magnetic field of quiescent prominences generally orientated horizontally (e.g. Mackay et al. 2020), we can connect the directions aligned with the flow, with the magnetic field and perpendicular to these two with the mixing layer model. This gives (connecting to equation 9) with $D_{\rm THREAD} = D_{\rm MIX}$ and $H_{\rm THREAD} = L_{\rm MIX}$. We also take that $\Delta V = V_{\rm DOWN}$ (i.e. the downflow velocity) and assume that a mixing layer can form on both sides of a downflow (see schematic a in Fig. 9). Therefore, we can adapt equation (9) to give the mass addition rate added to the cool phase, i.e.

$$\dot{M}_{\rm ADD} \approx \sqrt{2} C_1 \rho_1 T_{\rm THREAD} H_{\rm THREAD} \frac{\rho_1^{1/4} \rho_2^{3/4}}{(\sqrt{\rho_1} + \sqrt{\rho_2})^2} V_{\rm DOWN}. \quad (24)$$

Now that the rate of material created by a downflow has been estimated, we need an estimate of the rate at which a downflow can extract material from a prominence. If we again consider a single downflow, we can estimate that the mass flux through any cross-section of the downflow would be given by

$$\dot{M}_{\rm DOWN} = \rho_2 V_{\rm DOWN} W_{\rm THREAD} D_{\rm THREAD}, \quad (25)$$

where $W_{\rm THREAD}$ is the observed width of a downflowing thread. If we take that any downflow is extracting mass from a prominence then

$$\frac{\dot{M}_{\rm ADD}}{\dot{M}_{\rm DOWN}} = \sqrt{2} C_1 \frac{\rho_1}{\rho_2} \frac{H_{\rm THREAD}}{W_{\rm THREAD}} \frac{\rho_1^{1/4} \rho_2^{3/4}}{(\sqrt{\rho_1} + \sqrt{\rho_2})^2}. \quad (26)$$

Here, the ratio is independent of the downflow velocity, so as long as conditions for turbulence to develop are satisfied the mass addition rate would be the same for any downflow. For an aspect ratio of $H_{\rm THREAD}/W_{\rm THREAD} = 10$ and taking $C_1 = 0.5$, we find that $\dot{M}_{\rm ADD}/\dot{M}_{\rm DOWN} \approx 0.02$, so 2 per cent of the mass lost in a single downflow is added back into the prominence. However, if we take that material falling down but has not reached the base of the prominence is still part of the prominence, then a downflow only leads to mass drainage when the material leaves the base of the prominence and not before. Therefore this small mass addition could become more significant if we consider all downflows and not those transporting mass out of a prominence at any given time. If we assume a uniform probability of a downflowing thread existing at a given height in a prominence at any given time, this means that the probability that mass is exiting the prominence can be estimated as $f = H_{\rm THREAD}/H_{\rm PROM}$ (with $H_{\rm PROM}$ the height of the prominence). This leads to

$$\frac{\dot{M}_{\rm ADD}}{\dot{M}_{\rm DOWN}} = \sqrt{2} C_1 \frac{\rho_1}{\rho_2} \frac{H_{\rm PROM}}{W_{\rm THREAD}} \frac{\rho_1^{1/4} \rho_2^{3/4}}{(\sqrt{\rho_1} + \sqrt{\rho_2})^2}. \quad (27)$$

Assuming an aspect ratio of $H_{\rm PROM}/W_{\rm THREAD} = 100$ (reasonable for a quiescent prominence larger than 30 Mm, see panel b of Fig. 1) this gives $\dot{M}_{\rm ADD}/\dot{M}_{\rm DOWN} \approx 0.18$, so the equivalent of approximately 18 per cent of the mass of cool material extracted by downflows is condensed into the prominence by mixing-induced cooling. This is potentially a significant contribution to the prominence mass budget. One qualifier that should be highlighted here is if the downflow interacts not with the corona but with prominence material then this mixing process will not be able to drive condensation and mass gain. More studies are needed of prominence downflows to understand how this process may work.

As a note on the observational consequences of this process, as the addition of mass to the cool phase comes through a cooling driven by turbulence, it would be expected that the intensity in intermediate temperature lines like Si IV and the non-thermal line width of the spectral line would increase. The Interface Region Imaging Spectrograph (IRIS) has the potential to make these observations.

### 4.2 Explaining the fading of spicules and prominence threads

There are a number of observations that show fading of emission in cool ($\sim 10^4$ K) spectral bands and the co-temporal appearance of enhanced emission in warmer spectral bands ($\sim 10^5$ K). Important examples of these observations would be in spicules (Beckers 1972; de Pontieu et al. 2007; Sterling et al. 2010; Pereira et al. 2014; Antolin et al. 2018) and solar prominence threads (Okamoto et al. 2015). Both spicules and prominence threads can be seen in panel (a) of Fig. 1. It has been hypothesized that the KH instability could be active during this change in emission (e.g Antolin et al. 2015, 2018; Hillier & Arregui 2019). The key property predicted for the mixing layer between the prominence/spicule material ($10^4$ K) and the corona ($10^6$ K) is that the characteristic temperature of the mixing layer will be $T_{\rm mix} = 10^5$ K (Hillier & Arregui 2019). This temperature is at the peak of the $L(T)$ curve for the solar atmosphere (Anzer & Heinzel 2000). Therefore a growing mixing layer as we model in this paper developing between either of these cool phases and the hot phase of the corona is likely to induce efficient radiative losses in the mixing layer. For schematic, see panel (b) of Fig. 9.

To apply the model presented in this paper to interaction of a prominence thread or a spicule with the corona, it is necessary to determine the appropriate values for the key parameters of the model. To determine $T_{\rm CUT}$, we note that the cooling function for radiative losses is expected to drastically drop at temperatures near $10^4$ K. This is because at these temperatures the radiative losses are approximately balanced by the radiative gains through photon absorption as the cool material (e.g Heinzel, Vial & Anzer 2014). Therefore, it is reasonable to use the temperature of $10^4$ K as the estimate of $T_{\rm CUT}$. Taking $D_{\rm MIX}$ to be the diameter of the prominence thread or spicule implies an appropriate characteristic value is $D_{\rm MIX} = 10^7$ cm. Taking $L_{\rm MIX}$ as the length of the mixing layer along the prominence thread or spicule implies an appropriate value is $L_{\rm MIX} = 10^8$ cm. The rms velocity of the turbulence should be calculated using a velocity difference that is approximately half the velocity amplitude of any oscillation (Hillier, Van Doorsselaere & Karampelas 2020), because the velocity amplitude is only a measure of the peak shear flow. This therefore introduces a factor of 1/2 making $\Delta V = V_{\rm AMP}/2$. However, as shown in the schematic in Fig. 9, it is natural for mixing layers to form on opposite sides of an oscillating tube (e.g. Antolin et al. 2015), because the back-and-forth motions of the tube create large shear in two opposite regions on the tube surface (Goossens et al. 2009). This leads to a factor of 2 increase, cancelling the factor of 1/2 related to the velocity amplitude. Therefore, the estimate of the energy losses is given by

$$\dot{E}_{\rm loss} \approx 10^{19} \left(\frac{C_1}{0.5}\right) \left(\frac{L_{\rm MIX}}{10^8 {\rm cm}}\right) \left(\frac{D_{\rm MIX}}{10^7 {\rm cm}}\right)$$
$$\times \left(\frac{V_{\rm AMP}}{10^6 {\rm cm\ s^{-1}}}\right) \left(\frac{P}{0.1 {\rm dyn\ cm^{-2}}}\right) {\rm erg\ s^{-1}}. \quad (28)$$

This corresponds to an enhancement in the radiative loss flux ($F$) of

$$F_{\text{loss}} \approx 10^4 \left(\frac{C_1}{0.5}\right) \left(\frac{V_{\text{AMP}}}{10^6 \text{cm s}^{-1}}\right) \\ \times \left(\frac{P}{0.1\text{dyn cm}^{-2}}\right) \text{erg cm}^2 \text{ s}^{-1}. \quad (29)$$

From the energy loss rate, we can then predict the cooling time-scale from a volume of size $L_{\text{MIX}} D_{\text{MIX}}^2$ which gives

$$\tau_{\text{loss}} = 100 \left(\frac{D_{\text{MIX}}}{10^7 \text{cm}}\right) \left(\frac{C_1}{0.5}\right)^{-1} \left(\frac{V_{\text{AMP}}}{10^6 \text{cm s}^{-1}}\right)^{-1} \text{s}. \quad (30)$$

This is much longer than the actual radiative loss time of the material in the mixing layer, but it is reflective of the physical process of the cooling being induced by the mixing of the two phases of material. The time-scale predicted here for the cooling is approximately the same as the observed time-scale for spicules or prominence threads to fade in observations of cool spectral lines and become brighter in warm spectral lines (Pereira et al. 2014; Okamoto et al. 2015; Samanta et al. 2019).

We can compare this energy loss through mixing-induced radiative losses to the radiative losses expected for the original volume before mixing occurs. As the cooler material, i.e. the prominence or spicule material, will be closer to a radiative equilibrium, that is the energy lost through emitting photons is matched by the energy gained through absorbing photons, at temperatures of 8000 K (e.g. Heinzel et al. 2014), then we can assume that there is net zero radiative loss from the cool, dense phase. For the corona, we have $L(T) = 10^{-22}$ erg cm$^3$ s$^{-1}$ and a number density of $n = 5 \times 10^8$ cm$^{-3}$ which gives an energy density loss of $2.5 \times 10^{-10}$ erg cm$^{-3}$ s$^{-1}$. In a volume of $L_{\text{MIX}} D_{\text{MIX}}^2$, we have $\dot{E}_{\text{ORIG-loss}} = 2.5 \times 10^{17}$ erg$^{-1}$, a factor of 40 smaller than the cooling rate associated with mixing-induced radiative losses. Naturally, this reduction in the losses leads to a cooling time that is a factor of 40 longer. Note that these calculated losses only hold for the situation of a growing mixing layer. If that growth is halted then the cooling process discussed here would likely halt too.

*4.2.1 Consideration of the intensity evolution in cool spectral lines*

The theory presented here makes the direct prediction that when cool and hot material mix, the mixing produces warm material at transition region densities that has enhanced radiative losses, and as such easily cools to produce more cool material (though still at transition region densities). This leads to the question: why do we see the fading of spicules and prominence threads in cool spectral lines if more cool material is being added during this process? The answer relates to the change in nature of the cool phase. Before it is brought into the mixing layer the cool phase can be seen as a compact region with a high density. However, once the material is brought into the mixing layer and then cools, the average density of the cool phase in the mixing layer is that of the warm phase from which it cools, that is to say approximately an order of magnitude smaller than the original cool, dense phase and is far more diffuse. This would likely have consequences for the observed emission in cool spectral lines.

For example, the H$\alpha$ intensity of a prominence at fixed temperature relates to the local pressure (or density) and number density of the first excited state ($n_2$, Heinzel, Gunár & Anzer 2015). In equilibrium conditions, the $n_2$ population is given by $n_2 = n_e^2/f(T, P)$ with $f(T, P)$ taking a value between 4 and 10 (Heinzel et al. 2015). The lower density of the cooler material in the mixing layer results in an $n_2$ population which is two orders of magnitude smaller than the original prominence thread/spicule. The intensity is related to the column integral of the $n_2$ population, so taking into account the larger region the diffuse material in the mixing layer occupies implies that the H$\alpha$ intensity of the mixing region will be at least one order of magnitude smaller than the original prominence thread/spicule.

Clearly spectral lines with different optical depths will respond differently to H$\alpha$. However, even in the Mg II H&K lines, which have much higher opacity (Heinzel et al. 2014), it can be expected that the intensity will be between a factor of 2 and an order of magnitude smaller for the low density, diffuse material created by mixing compared with the original dense structure (Heinzel et al. 2014). Therefore, even with cool material remaining after the mixing process, its diffuse nature will mean that the intensities observed will be significantly reduced, providing an explanation for the observed fading from cool spectral lines of spicules (Pereira et al. 2014) and prominence threads (Okamoto et al. 2015).

## 5 A DISCUSSION ON THE ROLE OF MAGNETIC FIELDS

### 5.1 Magnetic fields and dynamics

Until this point, we have only focused on the benefits of the presence of strong magnetic fields, i.e. that the total pressure is only changed slightly by the loss of gas pressure through cooling, making the dynamics easier to model. However, even in the linear instability magnetic tension is known to work to suppress the growth of the instability (Chandrasekhar 1961). Therefore, it is important to discuss where the effects of the magnetic field might change the mixing processes discussed above.

To develop KH turbulence in the presence of magnetic fields, there are two conditions that need to be satisfied. These are that the system is linearly unstable (e.g. Soler et al. 2010) and the flow has sufficient energy to develop overturning vortices instead of the instability being non-linearly saturated by the magnetic field (Hillier 2019, 2020). It has been shown in simulations of both impulsively excited and driven MHD kink waves for conditions relevant for studying the solar atmosphere that wave amplitudes consistent with observations can develop KH turbulence even in the presence of strong magnetic fields (e.g. Terradas et al. 2008; Antolin et al. 2015, 2018; Magyar & Van Doorsselaere 2016; Howson et al. 2017; Karampelas, Van Doorsselaere & Antolin 2017; Karampelas, Van Doorsselaere & Guo 2019). This occurs both with and without the presence of a resonant layer in the initial conditions (Terradas et al. 2008; Magyar & Van Doorsselaere 2016). Therefore, we can expect that KH turbulence occurs in the solar atmosphere. However, can a statement on the expected regime of the mixing be made based on physical field strengths.

For our assumption that the field and the flow are perpendicular to each other, a non-linear criterion for the turbulence having sufficient energy to twist up the magnetic field leading to 3D-like evolution was given in cgs units by Hillier & Arregui (2019) as

$$V_A < \frac{2LV_{\text{turb,rms}}}{w} \quad \to \quad B < \frac{\sqrt{2\pi}L\Delta V}{w} \frac{(\rho_1 \rho_2)^{1/2}}{(\sqrt{\rho_1} + \sqrt{\rho_2})}, \quad (31)$$

with $L$ the half-length of the structure along the magnetic field, $w$ a given width of a mixing layer, $V_A$ the characteristic Alfvén speed of the layer given as $B/\sqrt{4\pi\sqrt{\rho_1\rho_2}}$, $B$ the magnetic field strength in Gauss, and $V_{\text{turb, rms}}$ the rms turbulent velocity as given by equation (33) of Hillier & Arregui (2019).

This inequality can be applied to both observations and simulations to see in which regime they may fall. For example, in the simulations of Magyar & Van Doorsselaere (2016), they use a flux tube mimicking a coronal loop with density $2.5 \times 10^{-15}$ g cm$^{-2}$ of radius $1.5 \times 10^8$ cm and half-length $0.6 \times 10^9$ cm, with a background medium with density $0.5 \times 10^{-15}$ g cm$^{-2}$. There largest perturbation amplitude for the velocity was $3.5 \times 10$ cm s$^{-1}$. Approximating $\Delta V$ by the velocity amplitude, $w$ by the radius and $L$ by the half length gives a critical $B = 5.4$ G. In those simulations, a field strength of 12.5 G is used, which is larger than this critical field strength implying a more 2D-like, or at least coherency developing along the direction of the magnetic field. In this case, a value of $C_1$ closer to 0.5 might be appropriate, but more investigations are necessary.

Looking at an active region prominence, we can take a width $w$ of $10^6$ cm, a half-length of a prominence thread of $L = 10^8$ cm (see Fig. 1), a $\Delta V = V_{\rm AMP}/2 = 5 \times 10^5$ cm s$^{-1}$, a coronal number density of $n = 5 \times 10^8$ cm$^{-3}$ and a density contrast of 100 gives an upper magnetic field strength of 6.5 G. As this number is relatively small for an active region, the magnetic field will likely be above this threshold implying a more 2D like evolution, dominated by large vortices. This would be similar to the structures seen in the simulations of Magyar & Van Doorsselaere (2016), making $C_1 \approx 0.5$ potentially the expected choice. However, as the magnetic field will be anchored in the photosphere far from the ends of the prominence thread, this would increase the effective $L$ value meaning there is still the possibility for structuring to develop along the field.

## 5.2 Magnetic fields and fluid mixing

Another area to consider for the role of magnetic fields is how the material at two different temperatures can come together when at the high magnetic Reynolds numbers (ratio of diffusion to advection time-scales) of either the cool or hot phases in the solar corona, the material is effectively tied to the magnetic field. Even though the thermal conduction across the magnetic field is significantly smaller than that along the magnetic field, the fractal structure of the mixing layer (Fielding et al. 2020) means that a significant volume of a mixing layer would have this small thermal conduction term active, with a small effect happening in many places making it more effective.

There are, however, processes that can enhance this process significantly. First, the turbulent structure of KH turbulence drives the formation of multiple regions of high current (Antolin et al. 2015) that can lead to magnetic reconnection. The reconnection process will result in material of different temperatures existing on the same field line allowing the parallel thermal conduction to efficiently transporting thermal energy throughout the mixing layer. This process was seen in the simulation of Botha, Arber & Hood (2011). The other option connects to the fact that the cool phase of the solar corona contains neutral material (e.g. Heinzel et al. 2015) which being without charge can slip across the magnetic field. Therefore, in the turbulent mixing layer, neutral particles move across the magnetic field throughout the mixing layer providing a process to turbulently transport thermal energy across the magnetic field (Hillier 2019).

## 5.3 Magnetic fields and mixing layer structure and evolution

The model presented in this paper is based on the self similar evolution of mixing layers as found in hydrodynamic flows (e.g. Winant & Browand 1974). However, radiative mixing layers have been studied in many different astrophysical systems (e.g. Begelman & Fabian 1990; Ji et al. 2019; Fielding et al. 2020), and there are some fundamental differences between those models and the one proposed in this paper. For those radiative mixing layers, the cooling inside the mixing layer is balanced by an enthalpy flux into the mixing layer (e.g. Fielding et al. 2020), which has not been invoked in the model in this paper. With the cooling dynamics, key aspects of the model including the thickness of the layer, and the distribution of the turbulence are highly sensitive to the ratio of the mixing to cooling (e.g. Tan, Oh & Gronke 2021). Again, this is not an aspect of the model proposed here, nor are these seen in the simulations presented in Section 3.2.

These changes in the mixing layer dynamics are a consequence of the key difference between the models discussed above and the one proposed in this paper, i.e. the presence of strong magnetic fields in our model (and in the systems to which it is applicable). In situations with weak magnetic fields (i.e. $\beta > 1$), the radiative cooling of the mixing layer removes a significant proportion of the total pressure. This will result in the mixing layer contracting to create a pressure equilibrium. With strong magnetic fields, the magnetic pressure dominates the total pressure removing the connection between the thermodynamics and dynamics allowing the model for the dynamical evolution to be independent of the cooling rate.

## 6 SUMMARY

In this paper, we have investigated the process by which mixing through non-linear KH instability turbulence can lead to radiative cooling and its consequences for the interaction between cool and hot material in the solar atmosphere. The key results are that we can derive reasonable approximations of the rate at which mass and thermal energy are brought into the mixing layer by turbulence, and the rate at which energy is subsequently lost through radiative cooling. Applying these results to prominences with many downflows shows that the downflows can drive mixing and this can add back into the prominence ∼18 per cent of the cool material lost by the downflows. This model can also explain the fading of spicules and prominence threads from cool spectral lines with the creation of emission in warm spectral lines over a time-scale of ∼100 s without any heating process. However, the model developed here is only for a growing mixing layer, and what might cause a mixing layer to stop growing and what happens after the layer stops growing require further work to understand both theoretically and in terms of the consequences for the solar atmosphere.

To summarize the concepts modelled in this paper, cooling through mixing requires the presence of both a KH-unstable flow at the interface between two temperature plasmas and radiative losses from the plasma. For this process to then produce increased, efficient radiative losses, the time-scale of the radiative losses in the mixing layer needs to be shorter than the mixing time-scale. Further, for the particular model presented in this paper to hold, then strong (i.e. low plasma $\beta$) magnetic fields approximately perpendicular to the flow direction are required. As discussed in this paper, there are clear instances where these conditions are satisfied in the solar atmosphere, and there are likely to be many other applications in different astrophysical systems.

## ACKNOWLEDGEMENTS

AH and BS are supported by STFC Research Grant No. ST/R000891/1 and ST/V000659/1. AH was also supported by his STFC Ernest Rutherford Fellowship grant number ST/L00397X/1. IA and AH are supported by project PID2021-127487NB-I00 from Ministerio de Ciencia e Innovación and FEDER funds. AH and BS would like to thank the Research Support Engineer team at the University of Exeter for the improvements they made to the (P̱IP) code which have been used in this study. AH would like to acknowledge Dr. Takeshi Matsumoto for the conversations and guidance on turbulent mixing. This work was supported by the Research Institute for Mathematical Sciences, an International Joint Usage/Research Center located in Kyoto University. This research was supported by the International Space Science Institute (ISSI) in Bern, through ISSI International Team project 457 'The Role of Partial Ionization in the Formation, Dynamics, and Stability of Solar Prominences'. Hinode is a Japanese mission developed and launched by ISAS/JAXA, with NAOJ as domestic partner and NASA and STFC (UK) as international partners. It is operated by these agencies in co-operation with ESA and NSC (Norway). This work used the DiRAC@Durham facility managed by the Institute for Computational Cosmology on behalf of the STFC DiRAC HPC Facility (www.dirac.ac.uk). The equipment was funded by BEIS capital funding via STFC capital grants ST/P002293/1, ST/R002371/1, and ST/S002502/1, Durham University and STFC operations grant ST/R000832/1. DiRAC is part of the National e-Infrastructure.

## DATA AVAILABILITY

The active branch of the (P̱IP) code is available on GitHub (https://github.com/AstroSnow/PIP). The specific version used in this manuscript is archived and available at https://doi.org/10.5281/zenodo.7002509. All simulation data is available upon reasonable request.

This paper has been typeset from a T<sub>E</sub>X/LAT<sub>E</sub>X file prepared by the author.